\documentstyle[aps,twocolumn,prb,epsf]{revtex}
\begin{document}
\title{Influence of $Cu$ on spin-polaron
tunneling in the ferromagnetic state of $La_{2/3}Ca_{1/3}Mn_{1-x}Cu_xO_3$
from the resistivity data}
\author{S. Sergeenkov$^{1,2}$, H. Bougrine$^{1,3}$, M. Ausloos$^{1}$, and
R. Cloots$^{4}$}
\address{$^{1}$SUPRAS, Institute of Physics, B5, University of Li$\grave e$ge,
B-4000 Li$\grave e$ge, Belgium\\
$^{2}$Bogoliubov Laboratory of Theoretical Physics,
Joint Institute for Nuclear Research,\\ 141980 Dubna, Moscow Region, Russia\\
$^{3}$SUPRAS, Montefiore Electricity Institute, B28, University of
Li$\grave e$ge, B-4000 Li$\grave e$ge, Belgium\\
$^{4}$SUPRAS, Institute of Chemistry, B6, University of Li$\grave e$ge,
B-4000 Li$\grave e$ge, Belgium\\}
\date{\today}
\draft
\maketitle
\begin{abstract}
Nearly a $50\%$ decrease of resistivity $\rho (T,x)$ (accompanied by a $5\%$
reduction of the peak temperature $T_C(x)$) due to just $4\%$ $Cu$
doping on the $Mn$ site of $La_{2/3}Ca_{1/3}Mn_{1-x}Cu_xO_3$ is observed.
Attributing the observed phenomenon to the substitution induced decrease of
the polaron energy $E_{\sigma }(x)$ below $T_C(x)$, all data are
found to be well fitted by the nonthermal coherent tunneling expression
$\rho (T,x)=\rho _0e^{-\gamma M^{2}(T,x)}$ assuming
$M(T,x)=M_R(x)+M_0(x)\tanh\left\{\sqrt{\left[T_C(x)/T\right]^{2}-1}\right\}$
for the magnetization in the ferromagnetic state. The best fits
through all the data points yield $M_0(x)\simeq \sqrt{1-x}M_0(0)$,
$M_R(x)\simeq \sqrt{x}M_0(0)$, and $E_{\sigma }(x)\simeq E_{\sigma }(0)
(1-x)^{4}$ for the $Cu$ induced modifications of the
$Mn$ spins dominated zero-temperature spontaneous magnetization, the
residual paramagnetic contribution, and spin-polaron tunneling energy,
respectively, with $E_{\sigma }(0)=0.12 eV$ and $2R\simeq10\AA$
for the spin-polaron's size.
\end{abstract}

\narrowtext

As is well known,~\cite{1} the ground state of the highly magnetoresistive
conductor $La_{2/3}Ca_{1/3}MnO_3$
is ferromagnetic (FM) and the paramagnetic-ferromagnetic transition is
accompanied by a sharp drop in resistivity below $T_C$. Such
a correlation is considered as a basic element for the so-called magnetically
induced electron localization scenario~\cite{2,3,4} in which the changes of
observable resistivity at low temperatures are related to the corresponding
changes of the local magnetization, and a coherent nonthermal tunneling
charge carrier transport mechanism dominates other diffusion processes.

The effects of elemental substitution on the properties of
$La_{2/3}Ca_{1/3}MnO_3$ have been widely studied in an attempt to further
shed some
light on the underlying transport mechanisms in this interesting
material.~\cite{5,6,7,8,9,10,11,12,13}
In particular, substitution of the rare-earth atoms (like $Y$ or $Gd$)
on the $La$ site leads to the lowering of the ferromagnetic (and
"metal-insulator")
transition temperature $T_C$ due mostly to the cation size
mismatch.~\cite{1,7,11,12,13}
At the same time, the reduction of $T_C$ and a rather substantial drop
of resistivity in the FM region due to $Mn$ ions replacement with
metals (like $Co$ or $Ni$) are ascribed to a weakening of the Zener
double-exchange interaction between two unlike ions.~\cite{1,13}
In other words, similar to the effects of an applied magnetic field,~\cite{4}
metal-ion doping was found~\cite{13} to decrease the polaron tunneling
energy barrier (thus increasing the correlation length).

In this paper we present a comparative study of
resistivity measurements on two manganite samples from the
$La_{2/3}Ca_{1/3}Mn_{1-x}Cu_xO_3$ family for $x=0$ and $x=0.04$ and for a wide
temperature interval (from $20K$ to $300K$).
As we shall see, this very small amount of impurity leads to a marked (factor
of two) drop in resistivity value, hardly understandable along the
conventional scattering theories. The data are in fact well fitted by a
nonthermal spin tunneling
expression for the resistivity assuming a magnetization $M(T,x)$ dependent
charge carrier correlation length $L(M)$.

The samples examined in this study were prepared by the standard solid-state
reaction from stoichiometric amounts of $La_{2}O_3$, $CaCO_3$, $MnO_2$, and
$CuO$ powders. The necessary heat treatment was performed in air, in alumina
crucibles at $1300^{\circ}C$ for 2 days to preserve the
right phase stoichiometry.
Powder X-ray diffraction patterns are characteristic of perovskites. No
appreciable changes in the diffraction patterns induced by $Cu$ doping have
been observed (suggesting thus that no structural changes have occured
after replacement of $Mn$ by $Cu$). Energy Dispersive X-ray microanalyses
confirm the presence of copper on the manganese crystallographic sites.

The electrical resistivity $\rho (T,x)$ was
measured using the conventional four-probe method. To avoid Joule and
Peltier effects, a dc current $I=1mA$ was injected (as a one second pulse)
successively on both sides of the sample. The voltage drop $V$ across the
sample was measured with high accuracy by a $KT256$ nanovoltmeter.
Fig.1 presents the temperature dependence of the resistivity
$\rho (T,x)$ for two $La_{2/3}Ca_{1/3}Mn_{1-x}Cu_xO_3$ samples, with $x=0$
and $x=0.04$, respectively. Notice a rather sharp (nearly a $50\%$) drop of
resistivity (both near the peak and on its low temperature side) for the doped
sample along with a small reduction of the transition (peak) temperature
$T_C(x)$ reaching $T_C(0)=265K$ and $T_C(0.04)=250K$, respectively.

Since no tangible structural changes have been observed upon copper doping,
the Jahn-Teller mechanism can be safely ruled out and the most reasonable
cause for the resistivity drop in the doped material is the reduction of
the spin-polaron tunneling energy $E_{\sigma }$ which
within the localization scenario~\cite{2,3,4}
is tantamount to an increase of the charge carrier correlation length
$L=\sqrt{2\hbar ^{2}/mE_{\sigma }}$ (here $m$ is an effective polaron mass).
In the FM region (below $T_C(x)$), the tunneling based resistivity
reads~\cite{2,3,4}
$\rho [M(T,x)]=\rho _se^{2R/L(M)}$
where $\rho _s^{-1}=e^{2}R^{2}\nu _{ph}N_m$ with
$R$ being the tunneling distance (and $2R$ being a size of a small spin
polaron),
$\nu _{ph}$ the phonon frequency, and $N_m$ the density of available states.
In turn, the correlation length $L(M)$ depends on the temperature and
concentration of copper $x$ through the corresponding dependencies of the
magnetization $M(T,x)$. Assuming, along the main lines of conventional
mean-field approximation schemes~\cite{3,4} that
$L(M)=L_0/(1-M^{2}/M_{L}^{2})$ (with $M_L$ being a fraction of the saturated
magnetization $M(0)$), we arrive at the following expression for the
tunneling dominated resistivity
\begin{equation}
\rho (T,x)=\rho _0e^{-\gamma M^{2}(T,x)},
\end{equation}
with $\rho _0=\rho _se^{2R/L_0}$ and $\gamma =2R/L_0M_L^{2}$.

To account for the observed behavior of the resistivity, we identify $T_C$
with the doping-dependent Curie temperature $T_C(x)$, and
consider that the temperature and $x$ dependence of the magnetization is
the sum of a classical Curie-Weiss contribution and a residual term,
namely
\begin{equation}
M(T,x)=M_R(x)+M_0(x)\tanh\left\{\sqrt{\left[T_C(x)/T\right]^{2}-1}\right\}.
\end{equation}
Specifically, $M_R(x)$ is interpreted as a $Cu$ induced
paramagnetic contribution while
$M_0(x)$ accounts for the deviation of the ferromagnetically aligned $Mn$
magnetic moments
of the undoped material in the presence of copper atoms.
In fact, save for the $M_R(x)$ term, Eq.(2) is an analytical (approximate)
solution of the well-known Curie-Weiss mean-field equation on spontaneous
magnetization, viz.
$M(T,x)/M(0,x)=\tanh\left\{\left[M(T,x)/M(0,x)\right](T_C(x)/T)\right\}$.
To exclude any extrinsic effects (like grain boundary scattering), we
consider the normalized resistivity
$\Delta \rho (T,x)/\Delta \rho (0,x)$ with $\Delta \rho (T,x)=\rho (T,x)-
\rho (T_C(x),x)$ and $\rho (0,x)$ being the resistivity taken at the lowest
available temperature. Fig.2 depicts the above-defined normalized resistivity
versus the reduced temperature $T/T_C(x)$ for the two samples. First of all,
notice that the $x=0$ and $x=0.04$ data merge both at low temperatures
and above $T_C(x)$, while starting from $T_C(x)$ and below the $Cu$-doped
(squares) and $Cu$-free (circles) samples follow different routes. On the
other hand, approaching $T_C(x)$ from low temperatures, a (most likely)
fluctuation driven~\cite{4} crossover from undoped to doped transport
mechanism is clearly seen near $T/T_C(x)\simeq 0.9$. The solid lines are
the best fits through all the data points according to Eqs.(1) and (2),
yielding
$M_0(0)/M_L=1.41\sqrt{L_0/2R}$, $M_R(0)=0$, $M_0(0.04)=0.98M_0(0)$, and
$M_R(0.04)=0.06M_0(0)$ for the model parameters.
Recalling that in our
present study the copper content is $x=0.04$, the above estimates can be cast
into the following explicit $x$ dependencies of the residual
$M_R(x)\simeq \sqrt{x}M_0(0)$ and spontaneous $M_0(x)\simeq \sqrt{1-x}M_0(0)$
magnetizations, giving rise to an exponential (rather than power) $x$
dependence of the observed resistivity $\rho (T,x)$ (see Eqs.(1) and (2)).
Furthermore, assuming (as usual~\cite{4}) $2R/L_0\simeq 1$ for the (undoped)
tunneling distance to correlation length ratio and using the found value of
the residual resistivity $\rho (T_C(0),0)=\rho _0\simeq 3.5m\Omega m$,
the density of states~\cite{2} $N_m\simeq 9\times 10^{26}m^{-3}eV^{-1}$
and the phonon frequency~\cite{4} $\nu _{ph}\simeq 2\times 10^{13}s^{-1}$
(estimated from Raman shift for optical $Mn-O$ modes), we obtain
$R\simeq 5\AA$ for an estimate of the tunneling distance (corresponding to
$2R\simeq 10\AA$ for a spin-polaron's size)
which in turn results in $L_0\simeq 10\AA$ (using a free electron
approximation for a polaron's mass $m$) and
$E_{\sigma }(0)\simeq 0.12 eV$ for a zero-temperature copper-free
carrier charge correlation length and the spin-polaron tunneling energy,
respectively, both in good agreement with
reported~\cite{1,2,3,4,7,9,13} estimates of these parameters in other systems.
Based on the above estimates, we can roughly estimate the copper
induced
variation of the correlation length $L(x)$ and the corresponding spin polaron
tunneling energy $E_{\sigma }(x)$. Indeed, according to the earlier
introduced definitions,
$L[M(T_C(x))]=L_0/(1-M_R^{2}(x)/M_L^{2})\simeq L_0/(1-2x)$ and
$E_{\sigma }(x)\propto L^{-2}(x)$ which lead to $L(x)\simeq L(0)/(1-x)^{2}$
and $E_{\sigma }(x)\simeq E_{\sigma }(0)(1-x)^{4}$
for small enough $x$. These explicit $x$ dependencies (along
with the composition variation of the transition temperature $T_C(x)\simeq
T_C(0)(1-x)$) remarkably correlate with the $Co$ induced changes in
$La_{2/3}Ca_{1/3}Mn_{1-x}Co_xO_3$ recently observed by
Rubinstein {\it et al.}~\cite{13}

Interestingly, the above-obtained estimates agree very well with the observed
peak (at $T_C(x)$) and residual (at $T\to 0$) resistivities defined through
the model parameters as follows (see Eqs.(1) and (2))
\begin{equation}
\rho (T_C(x),x)=\rho _0e^{-\gamma M_R^{2}(x)},
\end{equation}
and
\begin{equation}
\rho (0,x)=\rho _0e^{-\gamma M^{2}(0,x)},
\end{equation}
with $\rho _0$, $M_R(x)$, and $M(0,x)$ defined earlier. This provides thus an
elegant self-consistency check for the employed interpretation.

In summary, a rather substantial drop in resistivity $\rho (T,x)$ of
$La_{2/3}Ca_{1/3}Mn_{1-x}Cu_xO_3$ sample upon just $4\%$ $Cu$ doping is
reported.
Along with lowering the Curie point $T_C(x)$, the copper substitution is
argued to add a small paramagnetic contribution $M_R(x)$ to the $Mn$ spins
dominated spontaneous magnetization $M$ of the undoped material leading
to a small decrease of the spin-polaron tunneling energy $E_{\sigma }(x)$.
However, due to the tunneling dominated carrier transport process, this small
amount of impurity
is sufficient for the drastical changes in resistivity absolute value across
the whole temperature range.
The temperature and $x$ dependencies of the observed resistivity was found to
be rather well fitted by a nonthermal
coherent tunneling of spin polarons with a heuristic expression for the
magnetization $M(T,x)$ in the ferromagnetic state (as an approximate analytic
solution of the mean-field Curie-Weiss equation), resulting in the
exponential (rather than linear) doping dependence of $\rho (T,x)$.

\acknowledgments

Part of this work has been financially supported by the Action de Recherche
Concert\'ees (ARC) 94-99/174.
S.S. thanks FNRS (Brussels) for some financial support.

\newpage

\begin{figure}[htb]
\epsfxsize=9cm
\centerline{\epsffile{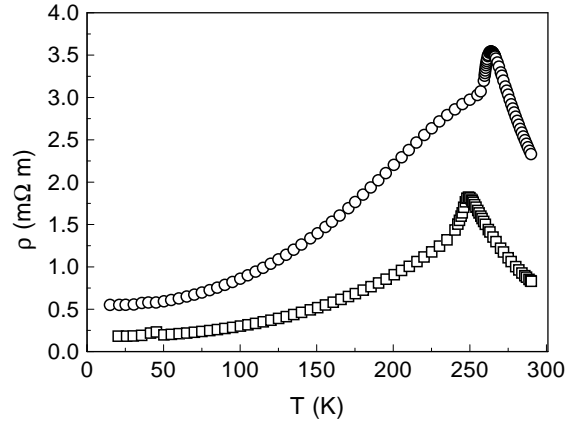} }
\caption{Temperature behavior of the observed resistivity
$\rho (T,x)$ in $La_{2/3}Ca_{1/3}Mn_{1-x}Cu_xO_3$
for $x=0$ (circles) and $x=0.04$ (squares).}
\end{figure}

\begin{figure}[htb]
\epsfxsize=9cm
\centerline{\epsffile{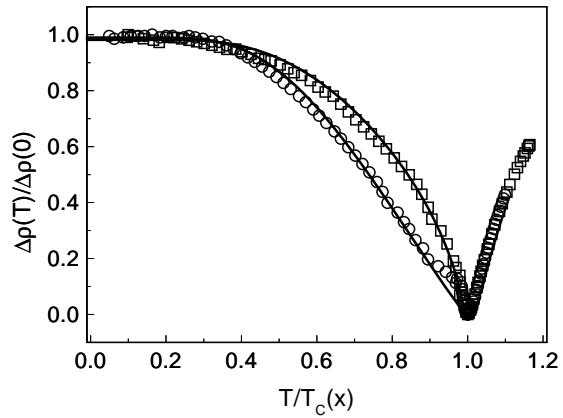} }
\caption{The temperature dependence of the normalized resistivity
$\Delta \rho (T,x)/\Delta \rho (0,x)$ in $La_{2/3}Ca_{1/3}Mn_{1-x}Cu_xO_3$
for $x=0$ (circles) and $x=0.04$ (squares) as a function of the reduced
temperature $T/T_C(x)$. The solid lines through all the data points are the
best fits according to Eqs.(1) and (2).}
\end{figure}

\end{document}